\documentclass[a4paper,fleqn]{cas-sc}
\usepackage{amsfonts,mathtools,amssymb,siunitx,mathrsfs,amsmath}
\usepackage{graphicx, nicefrac}
\usepackage{paralist}
\usepackage[numbers,sort&compress]{natbib}
\usepackage{scalerel}
\newcommand{\str}{\mathrm{Str}}

\usepackage{braket}

\usepackage{slashed}
\usepackage{cleveref}
\numberwithin{equation}{section}
\usepackage[colorinlistoftodos]{todonotes}
\usepackage{float}
\usepackage{indentfirst}
\usepackage{soul}
\usepackage{booktabs}
\usepackage{eso-pic,graphicx}
\usepackage{nicefrac}
\usepackage{epsfig}

\DeclareMathOperator\sign{sign}

\def\tsc#1{\csdef{#1}{\textsc{\lowercase{#1}}\xspace}}
\tsc{WGM}
\tsc{QE}
\tsc{EP}
\tsc{PMS}
\tsc{BEC}
\tsc{DE}

\begin{document}
\let\WriteBookmarks\relax
\def\floatpagepagefraction{1}
\def\textpagefraction{.001}
\shorttitle{Quantum integrability of a massive anisotropic SU(N) fermionic model}
\shortauthors{A. Melikyan, G. Weber.}

\title [mode = title]{Quantum integrability of massive anisotropic SU(N) fermionic models}
                      
\author[1]{\color{black}A. Melikyan}
\cormark[1]
\ead{amelik@gmail.com}
\address[1]{Instituto de F\'isica, Universidade de Bras\'ilia, 70910-900, Bras\'ilia, DF, Brasil}

\author[2]{\color{black}G. Weber}
\cormark[2]
\address[2]{Escola de Engenharia de Lorena, Universidade de S\~ao Paulo, 12602-810, Lorena, SP, Brazil}
\ead{gabrielleweber@usp.br}

\begin{abstract}
We consider a general anisotropic massive SU(N) fermionic model, and investigate its quantum integrability. In particular, by regularizing singular operator products, we derive a system of equations resulting in the S-matrix and find some non-trivial solutions. We illustrate our findings on the example of a SU(3) model, and show that the Yang-Baxter equation is satisfied in the massless limit for all coupling constants, while in the massive case the solutions are parameterized in terms of the exceptional solutions to the eight-vertex model.
\end{abstract}

\begin{keywords}
\end{keywords}
\maketitle

\section{Introduction: $SU(N)$ anisotropic model}
\label{ani}

The problem we set in this paper is to investigate the quantum integrability of a general massive anisotropic fermionic $SU(N)$ model, defined by the following Lagrangian:
\begin{align}
    L=i \bar{\psi} \gamma^{\mu} \partial_{\mu} \psi -m \bar{\psi} \psi -\frac{1}{2} g_{0} (\bar{\psi} \gamma^{\mu} \psi)^{2} 
    - \frac{1}{2} M_{AB} (\bar{\psi} \gamma^{\mu} T^{A} \psi)(\bar{\psi} \gamma_{\mu} T^{B} \psi). \label{ani:general_su_n_lagrangian}
\end{align}
Here, $g_{0}$ and $M_{AB}$, $A,B=1, \ldots, (N^2 -1)$ represent an array of coupling constants, and $T^{A}$, $A=1, \ldots, (N^2 -1)$, the $SU(N)$ generators in the following representation:
\begin{align}
    \left(T^{(1)}_{ab}\right)_{ij}=\frac{1}{2} \left[ \delta_{ai} \delta_{bj} +\delta_{aj} \delta_{bi} \right], \,
    \left(T^{(2)}_{ab}\right)_{ij}=-\frac{i}{2} \left[ \delta_{ai} \delta_{bj} -\delta_{aj} \delta_{bi} \right]; \quad  1 \le a < b \le N, \; 1\le i,j \le N \label{ani:T1_T2_generators}
\end{align}
and:
\begin{align}\label{ani:T3_generator}
   \left(T^{(3)}_{ab}\right)_{ij} = 
    \begin{cases}
      \omega(a) \delta_{ij}; &i<a \\
      -\tilde{\omega}(a) \delta_{ij}; &i=a \\
     0; &i>a
    \end{cases};\quad 2 \le a \le N, 1\le i,j \le N,
\end{align}
where we have denoted $\omega(a):= \left[ 2a(a-1) \right]^{\nicefrac{1}{2}}$ and $\tilde{\omega}(a):= \left[(a-1)/(2a)\right]^{\nicefrac{1}{2}}$. A simpler $SU(2)$ case, i.e., the Lagrangian \eqref{ani:general_su_n_lagrangian} for $N=2$ and a particular choice of the coupling constants, has already been considered in \cite{Belavin:1979pq,Dutyshev:1980vn} for the massless case and, more recently in \cite{Melikyan:2019ac}, for the massive case, where it was shown that the $2$-particle $S$-matrix has, in general, the form of the $R$-matrix for the inhomogeneous eight-vertex model \cite{Baxter:1982zz,Khachatryan:2012wy,Hietarinta:1991ti}. This allowed us to find some exceptional solutions to the Yang-Baxter equation, that turned out to be related to the $XXZ$ model. Our motivation to start from a massive theory is twofold: first, to obtain an effective theory in the low energy limit as a typical $\nicefrac{1}{m}$ expansion, which could be a candidate to match a highly non-trivial integrable fermionic model that was originally obtained while investigating strings on the $AdS_5 \times S^5$ background \cite{Beisert:2010jr,Alday:2005jm, Melikyan:2011uf}. We briefly note here that this model contains interaction terms up to the sixth order in fermion fields and their derivatives, and is considerably more complex than the massive Thirring model. The second, more important reason to start with a massive theory is the result originally obtained in \cite{Vaks:1961vl}, which states that even by starting with a zero bare mass in an anisotropic fermionic theory, a non-zero mass term is regenerated due to renormalization effects. Thus, our analysis is general from the beginning, in that we start from a massive theory, and the massless limit is considered later in the paper as one of possible cases. This is unlike the earlier works (e.g., \cite{Belavin:1979pq,Dutyshev:1980vn,Doria:1984ym,Reshetikhin:1985ru}) where only the massless cases were considered.\footnote{The problem regarding the analyticity of the $m \to 0$ limit, i.e., whether starting with a massive theory and taking $m \to 0$ limit one ends up with the same  results as starting with the massless theory will be considered elsewhere.}

It is convenient to cast the Lagrangian 
\eqref{ani:general_su_n_lagrangian} in the following form:
\begin{align}
    L=i \psi^{\dagger \alpha}_{i} \partial_{0} \psi^{\alpha}_{i}-i \psi^{\dagger \alpha}_{i_{1}} (\sigma^{3})_{i_{1}i_{2}}\partial_{1} \psi^{\alpha}_{i_{2}} 
    -m \psi^{\dagger \alpha}_{i_{1}}(\sigma^{1})_{i_{1}i_{2}} \psi^{\alpha}_{i_{2}} +\frac{1}{2}\psi^{\dagger \alpha_{1}}_{i_{1}}\psi^{\dagger \beta_{1}}_{j_{1}} 
    G_{\alpha_{1} \alpha_{2},\beta_{1} \beta_{2}}\Lambda_{i_{1} i_{2},j_{1} j_{2}}\psi^{\alpha_{2}}_{i_{2}}\psi^{\beta_{2}}_{j_{2}}, \label{ani:general_su_n_lagrangian_GL_tensors}
\end{align}
where we used $i,j = 1,2$ for spinor indices and $\alpha, \beta = 1, \ldots N$ for the isotopic ones. The $G$ and $\Lambda$ tensors in \eqref{ani:general_su_n_lagrangian_GL_tensors} are defined as follows:
\begin{align}
    G_{\alpha_{1} \alpha_{2},\beta_{1} \beta_{2}} &:= g_{0} \delta_{\alpha_{1} \alpha_{2}} \delta_{\beta_{1} \beta_{2}} + M_{AB} (T^{A})_{\alpha_{1} \alpha_{2}}
    (T^{B})_{\beta_{1} \beta_{2}}, \label{ani:G_tensor}\\
    \Lambda_{i_{1} i_{2},j_{1} j_{2}} &:=\delta_{i_{1} i_{2}} \delta_{j_{1} j_{2}} - (\sigma^{3})_{i_{1} i_{2}} (\sigma^{3})_{j_{1} j_{2}}.\label{ani:Lambda_tensor}
\end{align}
To quantize the model and obtain the spectrum of the theory \eqref{ani:general_su_n_lagrangian_GL_tensors}, it is necessary, as shown in \cite{Melikyan:2016gkd,Melikyan:2019ac}, to consider operator-valued distributions and regularize singular expressions involving operator products at the same point by means of a procedure involving the simultaneous regularization of the fields and operator products. We briefly recall here the mains definitions. The fields are treated as operator valued distributions,
\begin{align}\label{app:skl_operator}
	{\psi_{\mathcal{F}}}_i^{\alpha} (x) = \int dz \: \mathcal{F}_{\mu}(x,z) \: \psi_i^{\alpha}(z),
\end{align}
with a symmetric test function satisfying  $\mathcal{F}_{\mu}(x,z) \xrightarrow{\mu \to 0} \zeta \delta (x-z)$.
Here, $\zeta$ is a $c$-number fixed so that the algebra of regularized fields be finite in the limit $x\to z$. The remaining singularities, related to the possible discontinuities of the wave-function $\phi^{i_1 \cdots i_N}_{\alpha_1 \cdots \alpha_N} (x_1,\ldots x_N)$ and its derivatives, are dealt with by smearing the product of operators \cite{Sklyanin:1988} over a $k$-dimensional hypercube of side $\Delta$ around the coinciding point $x$ with all possible singular $(k-1)$-dimensional hyperplanes removed, 
\begin{align}\label{app:skl_product}
	{\psi_{\mathcal{F}}}_1(x) \circ \cdots \circ {\psi_{\mathcal{F}}}_k(x) := \lim_{\Delta V_i \to 0} \frac{1}{\sum_{i=1}^{k!}} \sum_{i=1}^{k!} \int_{\Delta V_i} d\xi_1 \cdots d\xi_k \: {\psi_{\mathcal{F}}}_1 \left(\xi_1 \right) \cdots {\psi_{\mathcal{F}}}_k\left(\xi_k \right). 
\end{align}
The resulting $k!$ disjoint regions of volume $\Delta V_i$ correspond to all possible orderings of the variables $\xi_1,\ldots, \xi_k$ separated by the length of a regularization parameter $\epsilon$. By combining \eqref{app:skl_operator} with \eqref{app:skl_product}, we can formulate a well-defined quantum theory for the Lagrangian \eqref{ani:general_su_n_lagrangian_GL_tensors} and construct the corresponding Hilbert space 
\begin{align}
	|\psi \rangle_N = \int dx_1 \cdots dx_N \: \phi^{i_1 \cdots i_N}_{\alpha_1 \cdots \alpha_N} (x_1,\ldots x_N) \: {\psi_{\mathcal{F}}}_{i_1}^{\alpha_1} (x_1) \cdots {\psi_{\mathcal{F}}}_{i_N}^{\alpha_N} (x_N) |0\rangle
\end{align}
over the pseudo-vacuum $|0\rangle$ following the steps outlined in \cite{Melikyan:2019ac}. Finally, it is worth noting that it is exactly this simultaneous regularization of both fields and operator products that allows one to exactly reproduce, in the classical limit, Maillet's symmetrization prescription for dealing with non-ultralocal classical integrable models \cite{Melikyan:2016gkd}.

\section{$S$-matrix}
\label{smat}
We start by giving one of our key results - the necessary condition that arises as a requirement to diagonalize in the $2$-particle sector the quantum Hamiltonian corresponding to 
\eqref{ani:general_su_n_lagrangian_GL_tensors}:
\begin{align}
    &\int  \psi^{\dagger \alpha_{1}}_{l_{1}} \psi^{\dagger \beta_{1}}_{l_{2}}\Gamma^{\alpha_{1} \beta_{1}}_{l_{1} l_{2}} dx =0, \\
    &\Gamma^{\alpha_{1} \beta_{1}}_{l_{1} l_{2}}:=i \delta^{\alpha_{1} \gamma_{1}} \delta^{\beta_{1} \gamma_{2}} (\sigma^{3})_{l_{1} k_{1}} \delta_{l_{2} k_{2}} {\Delta^{(-)}}^{k_{1} k_{2}}_{\gamma_{1} \gamma_{2}}
    + \xi(\zeta) G_{\alpha_{1} \gamma_{1},\beta_{1} \gamma_{2}} \Lambda_{l_{1} k_{1},l_{2} k_{2}} {\Delta^{(+)}}^{k_{1} k_{2}}_{\gamma_{1} \gamma_{2}}. \label{ani:Gamma_def_condition}
\end{align}
Here $\xi(\zeta)$ is a function of the fundamental parameter $\zeta$, which arises a result of the regularization of singular operator products,\footnote{It can be shown that the appearance of the function $\xi(\zeta)$ results in the renormalization of the coupling constants.} and we have also combined the regularized $2$-particle wave functions as follows:
\begin{align}
    {\Delta^{(\pm)}}^{k_{1} k_{2}}_{\gamma_{1} \gamma_{2}}:={\phi}^{k_{1} k_{2}}_{\gamma_{1} \gamma_{2}}(x+\epsilon,x) \pm {\phi}^{k_{1} k_{2}}_{\gamma_{1} \gamma_{2}}(x-\epsilon,x). \label{ani:Delta_pm} 
\end{align}
Using the definitions \eqref{ani:G_tensor} and \eqref{ani:Lambda_tensor}, we find the following system of linear equations for ${\Delta^{(\pm)}}^{k_{1} k_{2}}_{\gamma_{1} \gamma_{2}}$:
\begin{align}
    &{\Delta^{(-)}}^{11}_{\alpha_{1} \beta_{1}} - {\Delta^{(-)}}^{11}_{\beta_{1} \alpha_{1}} =0,\quad {\Delta^{(-)}}^{22}_{\alpha_{1} \beta_{1}} - {\Delta^{(-)}}^{22}_{\beta_{1} \alpha_{1}} =0,\nonumber\\
    &i \left[{\Delta^{(-)}}^{12}_{\alpha_{1} \beta_{1}} + {\Delta^{(-)}}^{21}_{\beta_{1} \alpha_{1}}\right] + 2\xi \left[G_{\alpha_{1} \gamma_{1},\beta_{1} \gamma_{2}} 
    {\Delta^{(+)}}^{12}_{\gamma_{1} \gamma_{2}} - G_{\beta_{1} \gamma_{1},\alpha_{1} \gamma_{2}} {\Delta^{(+)}}^{21}_{\gamma_{1} \gamma_{2}}\right]=0.\label{ani:system_1}
\end{align}
The set of  equations \eqref{ani:system_1} plays a central role in the determination of the 2-particle $S$-matrix and the subsequent analysis of the $S$-matrix factorization for $n \geq 3$ \cite{Baxter:1982zz,Zamolodchikov:1978xm, Belavin:1979pq,Dutyshev:1980vn,Samaj:2013yva}. To this end, we consider the standard Bethe Ansatz for the $n$-particle wave-function in the $P$-sector \cite{Korepin:1997bk,Samaj:2013yva,Dutyshev:1980vn}:\footnote{Here, $I=(1,2, \ldots, n)$ with $x_{1} < x_{2}< \ldots <x_{n}$ denotes the fundamental ordering, $P=(P_{1},P_{2}, \ldots,  P_{n})$ with  $x_{P_{1}}<x_{P_{2}}< \ldots,<x_{P_{n}}$, an arbitrary ordering sector, and $PQ$ stands for the product of permutations. Moreover, the $n$-particle wave-functions satisfy the anti-symmetry condition: $\phi^{i_{q_1} \cdots i_{q_n}}_{\alpha_{q_1}\cdots \alpha_{q_n}} (x_{q_1}, \ldots, x_{q_n}) = (-1)^{\sign(q) } \phi_{\alpha_1 \cdots \alpha_n}^{i_1 \cdots i_n}(x_1,\ldots, x_N).$}
\begin{align}
	\phi^{P}_{\alpha_{1}\cdots \alpha_{n}}(x) = \sum\limits_{Q}\: \sign(Q)\, A^{PQ}_{\alpha_{P_{1}}, \ldots, \alpha_{P_{n}}}\prod\limits_{j=1}^{n} u_{j}\left(\theta_{Q_{j}}\right) \exp\left(i k_{Q_{j}}x_{j}\right),\label{ani:wave_function}
\end{align}
where $k_{i}=m\sinh(\theta_{i})$, and $u_{j}(\theta)=\frac{1}{\sqrt{2 \cosh(\theta)}}\begin{pmatrix}
           e^{-\nicefrac{\theta}{2}} \\
           e^{\nicefrac{\theta}{2}} \\
		   \end{pmatrix}$ 
is the spinor for the  $j{\textrm{\small th}}$ particle. The quantum integrability of the model, expressed as the factorization of the $n \ge 3$-particle  $S$-matrix in terms of $2$-particle $S$-matrices, follows from \eqref{ani:wave_function}, provided the Yang-Baxter equation ($YBE$) is satisfied:
\begin{align}\label{ani:YBE}
	S_{12}(k_{1},k_{2})S_{13}(k_{1},k_{3})S_{23}(k_{2},k_{3})=S_{23}(k_{2},k_{3})S_{13}(k_{1},k_{3})S_{12}(k_{1},k_{2}),
\end{align}
where the $2$-particle $S$-matrix is defined via the exchange relation 
\begin{align} \label{ani:S_matrix_exchange}
\mathcal{A}_{\sigma_j \sigma_{i}}(k_{r},k_{s})=\sum\limits_{\sigma'_{i} \sigma'_{j}} S^{\sigma_{i} \sigma_{j}}_{\sigma'_{i} \sigma'_{j}}\mathcal{A}_{\sigma'_i \sigma'_{j}}(k_{s},k_{r}), 
\end{align}
with the $\mathcal{A}$-coefficients given by:
\begin{align}
\mathcal{A}_{\sigma_{P_{1}}\sigma_{P_{2}} \cdots \sigma_{P_{n}}}(k_{Q_{1}},k_{Q_{2}},\ldots,k_{Q_{n}})=A^{Q}_{\alpha_{P_{1}},\ldots, \alpha_{P_{n}}}\prod\limits_{j=1}^{n} u^{m_{P_{j}}}\left(\theta_{Q_{j}}\right).
\end{align}


Thus, upon substitution of \eqref{ani:wave_function} into the set of equations \eqref{ani:system_1}, one obtains a system of linear equations for the $\mathcal{A}_{\sigma_j \sigma_{i}}(k_{r},k_{s})$-amplitudes, from which one, in general, may determine the 2-particle $S$-matrix. The key problem is then to analyze the $YBE$
\eqref{ani:YBE} for the resulting $S$-matrix. For the $SU(2)$ fermionic models, as mentioned earlier, this was investigated in \cite{Belavin:1979pq,Dutyshev:1980vn} for the massless, and in \cite{Melikyan:2019ac} for the massive cases.\footnote{To make a connection with the notations of \cite{Melikyan:2019ac}, one should set: $g_{3}=\frac{1}{4}h^{(3)}_{2,2}; g_{4}=h^{(4)}_{2}=\tilde{h}^{(4)}_{2}; g_{1}-g_{2} =h^{(-)}_{12,12} ; g_{1}+g_{2}=h^{(+)}_{12,12}$, where $h^{(\pm)}_{\alpha \beta,\gamma \delta}$ are defined below \eqref{su3:system_lineq}.} In the latter case, it was shown that the above procedure results in the $2$-particle $S$-matrix, which has, in general, the form of the $R$-matrix for the inhomogeneous eight-vertex model \cite{Baxter:1982zz,Khachatryan:2012wy,Hietarinta:1991ti}, which allows some exceptional solutions to the $YBE$. 

For the $SU(N);\, N \ge 3$, the $YBE$ has a different structure in comparison to the $SU(2)$ case, leading to a substantially more complex analysis. To illustrate the key new points arising in this analysis, while avoiding unnecessary computational difficulties, we fix the coupling constants $g_{0}$ and $M_{AB}$ to consider the model corresponding to the following $G$ tensor \eqref{ani:G_tensor}:
\begin{align}
    G_{\alpha_{1} \alpha_{2},\beta_{1} \beta_{2}} &= g_{0}\delta_{\alpha_{1} \alpha_{2}} \delta_{\beta_{1} \beta_{2}} 
    + h^{(1)}_{ab,cd} \left(T^{(1)}_{ab}\right)_{\alpha_{1} \alpha_{2}} \left(T^{(1)}_{cd}\right)_{\beta_{1} \beta_{2}} 
    + h^{(2)}_{ab,cd} \left(T^{(2)}_{ab}\right)_{\alpha_{1} \alpha_{2}}\left(T^{(2)}_{cd}\right)_{\beta_{1} \beta_{2}} \nonumber \\
    &+ h^{(3)}_{a,b} \Big(T^{(3)}_{a}\Big)_{\alpha_{1} \alpha_{2}}\left(T^{(3)}_{b}\right)_{\beta_{1} \beta_{2}} + h^{(4)}_{a} \delta_{\alpha_{1} \alpha_{2}}\Big(T^{(3)}_{a}\Big)_{\beta_{1} \beta_{2}} + \tilde{h}^{(4)}_{a} \Big(T^{(3)}_{a}\Big)_{\alpha_{1} \alpha_{2}}\delta_{\beta_{1} \beta_{2}}. \label{ani:G_tensor_diag_choice}
\end{align}
Although much of the subsequent analysis can be performed in general for any $N$, in what follows, we also restrict ourselves to the $SU(3)$ case as a characteristic example which brings forward new features in the analysis of exact solutions. The system of equations \eqref{ani:system_1} for $N=3$ reduces to the following system of linear equations for the variables $z_{\alpha \beta}={\Delta^{(-)}}^{12}_{\alpha \beta} + {\Delta^{(-)}}^{21}_{\beta \alpha}$ and $y_{\alpha \beta}={\Delta^{(+)}}^{12}_{\alpha \beta} - {\Delta^{(+)}}^{21}_{\beta \alpha}$:\footnote{We stress that it is straightforward to repeat the above steps and find a similar system for the general $SU(N)$ case.}
\begin{align}
    i z_{11} +2\xi \left[ \eta_{1} y_{11} + h^{(-)}_{12,12} y_{22} + h^{(-)}_{12,13} y_{23} + h^{(-)}_{12,13} y_{32} 
    +h^{(-)}_{13,13} y_{33}\right]=0,\nonumber\\
    i z_{12} +2\xi \left[ \eta_{2} y_{12} + h^{(+)}_{12,12} y_{21} + h^{(-)}_{12,23} y_{23} + h^{(+)}_{12,13} y_{31} 
    +h^{(-)}_{13,23} y_{33}\right]=0,\nonumber\\
    i z_{21} +2\xi \left[ \eta_{2} y_{21} + h^{(+)}_{12,12} y_{12} + h^{(+)}_{12,13} y_{13} + h^{(-)}_{12,23} y_{32} 
    +h^{(-)}_{13,23} y_{33}\right]=0,\nonumber\\
    i z_{22} +2\xi \left[ \eta_{4} y_{22} + h^{(-)}_{12,12} y_{11} + h^{(+)}_{12,23} y_{13} + h^{(+)}_{12,23} y_{31} 
    +h^{(-)}_{23,23} y_{33}\right]=0,\nonumber\\
    i z_{13} +2\xi \left[ \eta_{5} y_{13} + h^{(+)}_{12,13} y_{21} + h^{(+)}_{12,23} y_{22} + h^{(+)}_{13,13} y_{31} 
    +h^{(+)}_{13,23} y_{32}\right]=0,\nonumber\\
    i z_{31} +2\xi \left[ \eta_{5} y_{31} + h^{(+)}_{12,13} y_{12} + h^{(+)}_{12,23} y_{22} + h^{(+)}_{13,13} y_{13} 
    +h^{(+)}_{13,23} y_{23}\right]=0,\nonumber\\
    i z_{33} +2\xi \left[ \eta_{9} y_{33} + h^{(-)}_{13,13} y_{11} + h^{(-)}_{12,23} y_{12} + h^{(-)}_{13,23} y_{21} 
    +h^{(-)}_{23,23} y_{22}\right]=0,\nonumber\\
    i z_{23} +2\xi \left[ \eta_{7} y_{23} + h^{(-)}_{12,13} y_{11} + h^{(-)}_{12,23} y_{12} + h^{(+)}_{13,23} y_{31} 
    +h^{(+)}_{23,23} y_{32}\right]=0,\nonumber\\
    i z_{32} +2\xi \left[ \eta_{7} y_{32} + h^{(-)}_{12,13} y_{11} + h^{(+)}_{13,23} y_{13} + h^{(-)}_{12,23} y_{32} 
    +h^{(+)}_{23,23} y_{23}\right]=0,\label{su3:system_lineq}
\end{align}
where we have denoted $h^{(\pm)}_{\alpha \beta, \gamma \delta}=\frac{1}{4}\left(h^{(1)}_{\alpha \beta, \gamma \delta}  \pm h^{(2)}_{\alpha \beta, \gamma \delta} \right)$. The constants $\eta_{1},\ldots,\eta_{9}$ in \eqref{su3:system_lineq} can be easily expressed as linear combinations of the original coupling constants (the explicit expressions are omitted here). Collecting the formulas \eqref{ani:Delta_pm} and \eqref{ani:wave_function}, and substituting them into \eqref{su3:system_lineq} one finally obtains the corresponding system of equations for $\mathcal{A}_{\sigma_j \sigma_{i}}(k_{r},k_{s})$. 

Even though the system \eqref{su3:system_lineq} can always be solved in general, the analytical form of the solution, depending on the model under consideration, i.e., the choice of the coupling constants, may be quite complex to deal with in practical calculation. For some cases, however, the above system can be decoupled into smaller subsets, making the analysis tractable. To give a concrete example, we further restrict the coupling constants by fixing $h^{(\pm)}_{ij,kl}=0$ for $\{i,j\} \neq \{k,l\}$. In this case, we can solve the exchange relation \eqref{ani:S_matrix_exchange} to obtain the following $S$-matrix:
\begin{align}
S(\theta) = \begin{pmatrix}
\alpha_{11}(\theta) & 0  & 0 & 0 & \alpha_{15}(\theta) & 0 & 0 & 0 & \alpha_{19}(\theta)\\
0 & \alpha_{22}(\theta) & 0 & \alpha_{24}(\theta) & 0 & 0 & 0 & 0 & 0 \\
0 & 0 & \alpha_{33}(\theta) & 0 & 0 & 0 & \alpha_{37}(\theta) & 0 & 0 \\
0 & \alpha_{24}(\theta) & 0 & \alpha_{22}(\theta) & 0 & 0 & 0 & 0 & 0 \\
\alpha_{15}(\theta) & 0  & 0 & 0 & \alpha_{55}(\theta) & 0 & 0 & 0 & \alpha_{59}(\theta)\\
0 & 0 & 0 & 0 & 0 & \alpha_{66}(\theta) & 0 & \alpha_{68}(\theta) & 0 \\
0 & 0 & \alpha_{37}(\theta) & 0 & 0 & 0 & \alpha_{33}(\theta) & 0 & 0 \\
0 & 0 & 0 & 0 & 0 & \alpha_{68}(\theta) & 0 & \alpha_{66}(\theta) & 0 \\
\alpha_{19}(\theta) & 0  & 0 & 0 & \alpha_{59}(\theta) & 0 & 0 & 0 & \alpha_{99}(\theta)
\end{pmatrix}. \label{smat:sm}
\end{align}
Even for this seemingly simple case, the resulting expressions are quite lengthy, and we relegate the detailed expressions for the non-zero $S$-matrix elements to Appendix \ref{app:smelem}. The $S$-matrix \eqref{smat:sm} can be checked to satisfy the standard normalization and unitary conditions. Moreover, it depends only on the difference of rapidities $\theta_{ij}:=\nicefrac{(\theta_{i}-\theta_{j})}{2}$.

\section{$YBE$}
\label{ybe}

Since the $S$-matrix \eqref{smat:sm} is of difference form, the $YBE$ \eqref{ani:YBE} takes the simpler form:
\begin{align}
\label{ani:YBE_rapidities}
S_{12}(\theta_{12})S_{13}(\theta_{13})S_{23}(\theta_{23})=S_{23}(\theta_{23})S_{13}(\theta_{13})S_{12}(\theta_{12}).
\end{align}
It corresponds to a system of 80 equations for the 12 independent non-zero S-matrix elements $\alpha_{ij}(\theta)$. We give the lengthy expressions in Appendix \ref{app:list_ybe}, and a brief look at Appendices \ref{app:smelem} and \ref{app:list_ybe} should suffice to convince a daunting task one is facing. Nevertheless, we can solve the $YBE$ in both massless limit and massive cases. The key point is to note that the $YBE$ in this case, unlike the simpler $SU(2)$ case which results in the usual eight-vertex model equations, has a nested structure with respect to the number of terms in each equation. Hence, we can split the system accordingly and, starting from the subset with the smallest number of terms, work our way up to recursively obtain constraints on the coupling constants. In the case of the example under consideration, the first equations in Appendix \ref{app:list_ybe}, all with two terms, yield the following relations:
\begin{align} 
	\frac{\alpha_{22}}{\alpha_{33}} = \kappa_{1}, \:
	\frac{\alpha_{22}}{\alpha_{66}} = \kappa_{2}, \:
	\frac{\alpha_{33}}{\alpha_{66}} = \kappa_{3}, \label{ybe:kappas}
\end{align}
where $\kappa_{i}$, $i=1,\ldots 3$ are constants. This set of equations, as can be easily seen by using the explicit formulas from Appendix \ref{app:list_ybe}, impose certain constraints on possible values of the coupling constants. Then, by substituting these relations in the remaining equations, we recursively eliminate variables and further fix the relations between the coupling constants. Although a lengthy and tedious procedure, it converges after a few steps. We postpone the detailed analysis to a future publication, and give here a summary for the massless limit and massive cases, the latter being much more complicated due to a non-trivial dependence on the rapidities via hyperbolic functions.

\emph{(I) Massless case:}
In order to take the limit $m \to 0$, we must first parametrize the rapidities as $\theta_n \to \sigma_n \tilde{\theta}$ and then let the parameter $\tilde{\theta} \to \infty$ simultaneously, so that the momenta $k_n = m \sinh \theta_n$ remain finite. The variable $\sigma_n$ corresponds to the helicity of the massless particles and takes the values $\pm 1$. In this limit, the relevant quantity $	\coth \theta_{ij} \to \sigma_{ij}, \: \text{with} \: \sigma_{ij} = \nicefrac{(\sigma_{i} - \sigma_{j})}{2}$ and the expressions for the non-zero $S$-matrix elements simplify considerably.

We now state one of the  main results of this paper: The $YBE$ is satisfied for all possible combinations of helicities without imposing any further constraint on the coupling constants. Namely, in the massless limit, the $S$-matrix \eqref{smat:sm} is a solution of the $YBE$ \eqref{ani:YBE_rapidities} for any $\eta_1, \ldots, \eta_9$ and any $h^{(\pm)}_{12,12}, h^{(\pm)}_{13,13}, h^{(\pm)}_{23,23}$. This result is analogous to the one considered by Dutyshev for the $SU(2)$ case in \cite{Dutyshev:1980vn}.\footnote{As a matter of fact, if we restrict our solution to the $SU(2)$ case, by setting
\begin{align}
	\eta_1 = \frac{\lambda_1+ \lambda_2}{4 \xi }, \: \eta_2 = \frac{\lambda_0 + \lambda_3}{4 \xi }, \: \eta_3 = \frac{\lambda_0 + \lambda_3}{4 \xi }, \: \eta_4 = \frac{\lambda_1+\lambda_2}{4 \xi }, \: h^{(-)}_{12,12} = \frac{\lambda_1-\lambda_2}{4 \xi }, \:  h^{(+)}_{12,12} = \frac{\lambda_3- \lambda_0}{4 \xi }
\end{align}
and all the other constants to zero, we obtain Dutyshev's $S$-matrix $S_D(\theta)$ (see \cite{Dutyshev:1980vn} for explicit expressions) as a direct sum decomposition of the form $S (\theta) = S_{D}(\theta) \oplus \mathbb{1}_5$, where $\mathbb{1}_n$ denotes the $n$-dimensional identity matrix.} The explicit expressions for the non-zero $S$-matrix elements in the massless limit are given at the end of Appendix \ref{app:smelem}. 

\emph{(II) Massive case:}
The massive case, as mentioned above, is considerably more complicated, given the structure of the non-zero $S$-matrix elements, which involve hyperbolic functions (see Appendix \ref{app:smelem}). Nonetheless, by using the nested structure of the $YBE$, as explained above, we obtain all possible solutions to the $YBE$. They correspond to $S$-matrices of the following possible forms:\footnote{We omit here solutions corresponding to complex coupling constants.}
\paragraph{\bf (1)} $S (\theta) = S_{M}(\theta) \oplus \mathbb{1}_5$, where $S_{M}(\theta)$, $M=1,2,3,4$ have the following form:
\begin{align}
	S_1(\theta) &=\left(
	\begin{array}{cccc}
	1 & 0 & 0 & 0 \\
	0 & \frac{\sinh (\theta ) \left(16 \xi ^2 g_0^2+1\right)}{\sinh (\theta )+8 \xi  g_0 \left(i \cosh (\theta )-2 \xi  \sinh (\theta ) g_0\right)} & \frac{8 i \xi  g_0}{\sinh (\theta )+8 \xi  g_0 \left(i \cosh (\theta )-2 \xi  \sinh (\theta ) g_0\right)} & 0 \\
	0 & \frac{8 i \xi  g_0}{\sinh (\theta )+8 \xi  g_0 \left(i \cosh (\theta )-2 \xi  \sinh (\theta ) g_0\right)} & \frac{\sinh (\theta ) \left(16 \xi ^2 g_0^2+1\right)}{\sinh (\theta )+8 \xi  g_0 \left(i \cosh (\theta )-2 \xi  \sinh (\theta ) g_0\right)} & 0 \\
	0 & 0 & 0 & 1 \\
	\end{array}
	\right),	
	\\
	S_2^{(\pm)}(\theta) &=
	\left(
	\begin{array}{cccc}
	\frac{1}{4 i \xi  g_0 \tanh \left(\frac{\theta }{2}\right)+1} & 0 & 0 & \pm \frac{4 i \xi  g_0}{\coth \left(\frac{\theta }{2}\right)+4 i \xi  g_0} \\
	0 & \frac{1}{4 i \xi  \coth \left(\frac{\theta }{2}\right) g_0+1} & \frac{4 \xi  g_0}{4 \xi  g_0-i \tanh \left(\frac{\theta }{2}\right)} & 0 \\
	0 & \frac{4 \xi  g_0}{4 \xi  g_0-i \tanh \left(\frac{\theta }{2}\right)} & \frac{1}{4 i \xi  \coth \left(\frac{\theta }{2}\right) g_0+1} & 0 \\
	\pm \frac{4 i \xi  g_0}{\coth \left(\frac{\theta }{2}\right)+4 i \xi  g_0} & 0 & 0 & \frac{1}{4 i \xi  g_0 \tanh \left(\frac{\theta }{2}\right)+1} \\
	\end{array}
	\right), 
	\\
	S_3^{(\pm)}(\theta) &= \frac{\mp \sqrt{16 g_0^2 \xi ^2+3} \pm 4 g_0 \xi +3 i \coth \left(\frac{\theta }{2}\right)}{\pm \sqrt{16 g_0^2 \xi ^2+3} \mp 4 g_0 \xi +3 i \coth \left(\frac{\theta }{2}\right)}
	\left(
	\begin{array}{cccc}
	1 & 0 & 0 & 0 \\
	0 & 0 & 1 & 0 \\
	0 & 1 & 0 & 0 \\
	0 & 0 & 0 & 1 \\
	\end{array}
	\right),
	\\
	S_4(\theta) &=S_T(\theta) \oplus \mathbb{1}_2 \oplus S_T(\theta),
\end{align}
where 
\begin{align}
	S_T(\theta) &=\frac{i + 4 g_0 \xi  \tanh \left(\frac{\theta }{2}\right)}{i-4 g_0 \xi  \tanh \left(\frac{\theta }{2}\right)}
\end{align}
is the $S$-matrix for the massive Thirring model \cite{Korepin:1997bk}. Thus, solution $S_4(\theta)$ corresponds to the decoupling of two copies of the massive Thirring model. The $S_{M}(\theta)$-matrices above are precisely all the massive exceptional solutions found in \cite{Melikyan:2019ac}, and, therefore, we have shown that, for the choice of coupling constants under consideration, the possible massive integrable models for the $SU(3)$ case are classified in terms of the massive exceptional solutions for the $SU(2)$ case. We also note that these $S_M(\theta)$ solutions appear for all the choices of coupling constants corresponding to the different $SU(2)$ sectors. 

\paragraph{\bf (2)} $S(\theta) = S_1(\theta) \oplus S_T(\theta) \oplus \mathbb{1}_4$: this solution appears in the following three cases 
\begin{inparaenum}[(i)]
	\item $\eta_1 = \eta_7 = 2 g_0$; 
	\item $\eta_4 = \eta_5 =2 g_0$; and
	\item $\eta_2 = \eta_9 = 2 g_0$
\end{inparaenum}
with all the other coupling constants set to zero. It non-trivially extends the $SU(2)$ solution corresponding to $S_1(\theta)$, originally derived in \cite{Melikyan:2019ac}, to the $SU(3)$ case. Noting that $S_T(\theta)$ is precisely the $S$-matrix for the massive Thirring model, it is clear that this solution corresponds to the decoupling of the massive anisotropic $SU(2)$ theory considered in \cite{Melikyan:2019ac} from the massive Thirring model \cite{Korepin:1997bk}. 

\paragraph{\bf (3)} $S(\theta) = S_T(\theta) \oplus \mathbb{1}_8$: this solution corresponds to the massive Thirring model.

\paragraph{\bf (4)} Trivial solutions corresponding to all the coupling constants fixed to a constant value inversely proportional to $\xi$. In these cases, all non-zero $S$-matrix elements are some simple combinations of hyperbolic functions of the rapidities. 

\paragraph{}

Although we have analyzed here only one possible solution to the system \eqref{su3:system_lineq} corresponding to our $SU(3)$ example, we have found other, more complex and interesting solutions that can be analyzed in an analogous manner. We stress that, despite the staggering number of equations to deal with, the key point is the fact that the $YBE$ contains also a subset of simpler equations (c.f. \eqref{ybe:kappas}), which immediately impose a number of constraints on the coupling constants, allowing us to simplify the remaining equations. In principle, it should be possible to repeat these steps and find some non-trivial solutions for some interesting $SU(N)$ anisotropic models in general. This more complete analysis will be presented in a future publication.

\appendix

\section{S-matrix elements}
\label{app:smelem}
In this appendix, we collect the non-zero $S$-matrix elements \eqref{smat:sm}:
\begin{small}
\begin{align}
\alpha_{11}(\theta) &= f_1\left(h_{12,12}^{(-)},h_{13,13}^{(-)},h_{23,23}^{(-)},\eta_1,\eta_4,\eta_9\right), \quad
\alpha_{55}(\theta) = f_1\left(h_{12,12}^{(-)},h_{23,23}^{(-)},h_{13,13}^{(-)},\eta_4,\eta_1,\eta_9\right), 
\\
\alpha_{99}(\theta) &= f_1\left(h_{13,13}^{(-)},h_{23,23}^{(-)},h_{12,12}^{(-)},\eta_9,\eta_1,\eta_4\right), \quad
\alpha_{15}(\theta) = f_2 \left(h_{12,12}^{(-)},h_{13,13}^{(-)},h_{23,23}^{(-)},\eta_9 \right), 
\\
\alpha_{19}(\theta) &= f_2 \left(h_{13,13}^{(-)},h_{23,23}^{(-)},h_{12,12}^{(-)},\eta_4 \right), \quad
\alpha_{59}(\theta) = f_2 \left(h_{23,23}^{(-)},h_{12,12}^{(-)},h_{13,13}^{(-)},\eta_1 \right), 
\\
\alpha_{22}(\theta) &= f_3\left(h_{12,12}^{(+)},\eta_2\right), \quad 
\alpha_{33}(\theta) = f_3\left(h_{13,13}^{(+)},\eta_5\right), \quad 
\alpha_{66}(\theta) = f_3\left(h_{23,23}^{(+)},\eta_7\right) 
\\
\alpha_{24}(\theta) &= f_4\left(h_{12,12}^{(+)},\eta_2\right), \quad 
\alpha_{37}(\theta) = f_4\left(h_{13,13}^{(+)},\eta_5\right), \quad 
\alpha_{68}(\theta) = f_4\left(h_{23,23}^{(+)},\eta_7\right).
\end{align}
\end{small}
In the formulas above, we introduced the functions:
\begin{small}
\begin{align}
f_1&(x_1,x_2,x_3,\eta_1,\eta_2,\eta_3) = \frac{1}{d_1} n_1(x_1,x_2,x_3,\eta_1,\eta_2,\eta_3), \\
f_2&(h_1,h_2,h_3,\eta)  = \frac{1}{d_2} n_2(x_1,x_2,x_3,\eta), \\
f_3&(h,\eta) = \frac{\sinh (\theta ) \left[1-4 \xi ^2 (h-\eta ) (\eta +h)\right]}{4 \xi ^2 (h-\eta ) (\eta +h) \sinh (\theta )-4 i \xi  (h-\eta  \cosh (\theta ))+\sinh (\theta )}, \\
f_4&(h,\eta) = \frac{4 i \xi  (\eta -h \cosh (\theta ))}{4 \xi ^2 (h-\eta ) (\eta +h) \sinh (\theta )-4 i \xi  (h-\eta  \cosh (\theta ))+\sinh (\theta )},
\end{align}
\end{small}
which are written in terms of
\begin{small}
\begin{align}
d_1 &= -2 \sinh ^3\left(\frac{\theta }{2}\right) \left\{ 16 \xi ^3 h_{12,12}^{(-)} h_{13,13}^{(-)} h_{23,23}^{(-)} +4 \xi ^2 h_{12,12}^{(-)}{}^2 \left(-2 \eta _9 \xi +i \coth \left(\frac{\theta }{2}\right)\right)+4 \xi ^2 h_{13,13}^{(-)}{}^2 \left(-2 \eta _4 \xi +i \coth \left(\frac{\theta }{2}\right)\right) \right. \nonumber \\
&+ \left. \left(2 \eta _1 \xi -i \coth \left(\frac{\theta }{2}\right)\right) \left[-4 \xi ^2 h_{23,23}^{(-)}{}^2-\left(\coth \left(\frac{\theta }{2}\right)+2 i \eta _4 \xi \right) \left(\coth \left(\frac{\theta }{2}\right)+2 i \eta _9 \xi \right)\right]\right\},
\\
d_2 &= -4 \left(\eta _1+\eta _4+\eta _9\right) \xi  \sinh ^2(\theta ) \sinh ^2\left(\frac{\theta }{2}\right) + 16 i \xi ^2 \sinh (\theta ) \sinh ^4\left(\frac{\theta }{2}\right) \left(-\eta _4 \eta _9-\eta _1 \left(\eta _4+\eta _9\right)+h_{12,12}^{(-)}{}^2+h_{13,13}^{(-)}{}^2+h_{23,23}^{(-)}{}^2\right) \nonumber \\ 
&+i \sinh ^3(\theta ) -64 \xi ^3 \sinh ^6\left(\frac{\theta }{2}\right) \left[h_{12,12}^{(-)} \left(\eta _9 h_{12,12}^{(-)}-2 h_{13,13}^{(-)} h_{23,23}^{(-)}\right)+\eta _4 h_{13,13}^{(-)}{}^2+\eta _1 \left(h_{23,23}^{(-)}{}^2-\eta _4 \eta _9\right)\right],
\\
n_1&(h_1,h_2,h_3,\eta_1, \eta_2, \eta_3) = 32 h_1 h_2 h_3 \xi ^3 \sinh ^3\left(\frac{\theta }{2}\right) + 4 i h_2^2 \xi ^2 \sinh \left(\frac{\theta }{2}\right) \left(\sinh (\theta )+2 i \eta _2 \xi  (\cosh (\theta )-1)\right)\nonumber \\
&+4 i h_1^2 \xi ^2 \sinh \left(\frac{\theta }{2}\right) \left(\sinh (\theta )+2 i \eta _3 \xi  (\cosh (\theta )-1)\right) \nonumber \\
&-i \left[\cosh \left(\frac{\theta }{2}\right)-2 i \eta _1 \xi  \sinh \left(\frac{\theta }{2}\right)\right] \left(2 i \left(\eta _2+\eta _3\right) \xi  \sinh (\theta )+4 \xi ^2 \left(h_3^2-\eta _2 \eta _3\right) (\cosh (\theta )-1)+\cosh (\theta )+1\right), 
\\
n_2&(h_1,h_2,h_3,\eta) = 16 \xi  \sinh ^4\left(\frac{\theta }{2}\right) \left[h_1 (2 i \eta  \xi  \sinh (\theta )+\cosh (\theta )+1)-2 i h_2 h_3 \xi  \sinh (\theta )\right].
\end{align}
\end{small}
In the massless limit, the relevant quantities for writing the non-zero $S$-matrix elements become:
\begin{small}
\begin{align}
	d_1&, d_2 \to +8 \xi ^3 h_{12,12}^{(-)} h_{13,13}^{(-)} h_{23,23}^{(-)} +2 \xi ^2 h_{12,12}^{(-)}{}^2 \left(-2 \eta _9 \xi +i \sigma \right) +2 \xi ^2 h_{13,13}^{(-)}{}^2 \left(-2 \eta _4 \xi +i \sigma \right)\nonumber \\ 
	&+ \frac{1}{2} \left(2 \eta _1 \xi -i \sigma \right) \left[-4 \xi ^2 h_{23,23}^{(-)}{}^2-\left(\sigma +2 i \eta _4 \xi \right) \left(\sigma +2 i \eta _9 \xi \right)\right],
	\\
	n_1&(h_1,h_2,h_3,\eta_1, \eta_2, \eta_3) \to -8 h_1 h_2 h_3 \xi ^3 + 2 h_2^2 \xi ^2 \left(2 \eta _2 \xi -i \sigma \right)+2 h_1^2 \xi ^2 \left(2 \eta _3 \xi -i \sigma \right)\nonumber \\
	&+\frac{1}{2} \left(2 \eta _1 \xi +i \sigma \right) \left[4 h_3^2 \xi ^2+\left(\sigma +2 i \eta _2 \xi \right) \left(\sigma +2 i \eta _3 \xi \right)\right]
	\\
	n_2&(h_1,h_2,h_3,\eta) \to 2 \xi  \sigma  \left[h_1 (\sigma +2 i \eta  \xi )-2 i h_2 h_3 \xi \right],
	\\
	f_3&(h,\eta) \to \frac{\sigma -4 \xi ^2 \sigma  (h-\eta ) (\eta +h)}{(2 \xi  (\eta +h)-i \sigma ) (2 \xi  \sigma  (h-\eta )+i)},
	\\
	f_4&(h,\eta) \to -\frac{2 i \xi  \left(\eta +\sigma ^2 (h-\eta )+h\right)}{(2 \xi  (\eta +h)-i \sigma ) (2 \xi  \sigma  (h-\eta )+i)}.
\end{align}
\end{small}

\section{List of Yang-Baxter Equations}
\label{app:list_ybe}

In this appendix, we explicitly display all the Yang-Baxter equations following from the $S$-matrix \eqref{smat:sm}. For the sake of clarity, we use the following shorthand notation: $\alpha_{ij}(\theta_{12}) = \alpha_{ij}$, $\alpha_{ij}(\theta_{13}) = \alpha_{ij}'$, $\alpha_{ij}(\theta_{23}) = \alpha_{ij}''$: 
\begin{small}
\begin{align}
\left(\alpha _{66} \alpha _{33}'-\alpha _{33} \alpha _{66}'\right) \alpha _{24}''=0, \quad 
\alpha _{68} \left(\alpha _{33}' \alpha _{22}''-\alpha _{22}' \alpha _{33}''\right)=0, \quad 
\alpha _{59}' \left(\alpha _{33} \alpha _{22}''-\alpha _{22} \alpha _{33}''\right)=0, 
\\
\left(\alpha _{66} \alpha _{22}'-\alpha _{22} \alpha _{66}'\right) \alpha _{37}''=0, \quad 
\alpha _{37} \left(\alpha _{66}' \alpha _{22}''-\alpha _{22}' \alpha _{66}''\right)=0, \quad 
\alpha _{24} \left(\alpha _{66}' \alpha _{33}''-\alpha _{33}' \alpha _{66}''\right)=0,
\\
\alpha _{51}' \left(\alpha _{66} \alpha _{33}''-\alpha _{33} \alpha _{66}''\right)=0, \quad 
\alpha _{91}' \left(\alpha _{66} \alpha _{22}''-\alpha _{22} \alpha _{66}''\right)=0, \quad 
\left(\alpha _{33} \alpha _{22}'-\alpha _{22} \alpha _{33}'\right) \alpha _{68}''=0, 
\\
\alpha _{24} \alpha _{37}' \alpha _{22}''+\alpha _{22} \alpha _{68}' \alpha _{24}''-\alpha _{68} \alpha _{22}' \alpha _{37}''=0, \quad
\alpha _{33} \alpha _{59}' \alpha _{24}''-\alpha _{51} \alpha _{91}' \alpha _{33}''-\alpha _{59} \alpha _{33}' \alpha _{37}''=0, 
\\
\alpha _{91} \alpha _{51}' \alpha _{22}''+\alpha _{59} \alpha _{22}' \alpha _{24}''-\alpha _{22} \alpha _{59}' \alpha _{37}''=0, \quad
\alpha _{68} \alpha _{33}' \alpha _{24}''-\alpha _{37} \alpha _{24}' \alpha _{33}''-\alpha _{33} \alpha _{68}' \alpha _{37}''=0, 
\\
\alpha _{24} \alpha _{59}' \alpha _{33}''-\alpha _{33} \alpha _{91}' \alpha _{51}''-\alpha _{37} \alpha _{33}' \alpha _{59}''=0, \quad
\alpha _{68} \alpha _{51}' \alpha _{33}''-\alpha _{37} \alpha _{33}' \alpha _{51}''-\alpha _{33} \alpha _{91}' \alpha _{59}''=0, 
\\
\alpha _{24} \alpha _{68}' \alpha _{22}''+\alpha _{22} \alpha _{37}' \alpha _{24}''-\alpha _{37} \alpha _{22}' \alpha _{68}''=0, \quad
\alpha _{24} \alpha _{66}' \alpha _{37}''-\alpha _{68} \alpha _{37}' \alpha _{66}''-\alpha _{66} \alpha _{24}' \alpha _{68}''=0, 
\\
\alpha _{37} \alpha _{68}' \alpha _{33}''+\alpha _{33} \alpha _{24}' \alpha _{37}''-\alpha _{24} \alpha _{33}' \alpha _{68}''=0, \quad
\alpha _{37} \alpha _{66}' \alpha _{24}''-\alpha _{68} \alpha _{24}' \alpha _{66}''-\alpha _{66} \alpha _{37}' \alpha _{68}''=0, 
\\
\alpha _{59} \alpha _{91}' \alpha _{33}''+\alpha _{51} \alpha _{33}' \alpha _{37}''-\alpha _{33} \alpha _{51}' \alpha _{68}''=0, \quad
\alpha _{66} \alpha _{51}' \alpha _{37}''-\alpha _{91} \alpha _{59}' \alpha _{66}''-\alpha _{51} \alpha _{66}' \alpha _{68}''=0, 
\\
\alpha _{66} \alpha _{91}' \alpha _{24}''-\alpha _{51} \alpha _{59}' \alpha _{66}''-\alpha _{91} \alpha _{66}' \alpha _{68}''=0, \quad
\alpha _{59} \alpha _{51}' \alpha _{22}''+\alpha _{91} \alpha _{22}' \alpha _{24}''-\alpha _{22} \alpha _{91}' \alpha _{68}''=0, 
\\
\alpha _{68} \alpha _{91}' \alpha _{22}''-\alpha _{22} \alpha _{51}' \alpha _{59}''-\alpha _{24} \alpha _{22}' \alpha _{91}''=0, \quad
\alpha _{37} \alpha _{59}' \alpha _{22}''-\alpha _{24} \alpha _{22}' \alpha _{59}''-\alpha _{22} \alpha _{51}' \alpha _{91}''=0, 
\\
\alpha _{68} \alpha _{66}' \alpha _{51}''-\alpha _{37} \alpha _{51}' \alpha _{66}''+\alpha _{66} \alpha _{59}' \alpha _{91}''=0, \quad
\alpha _{66} \alpha _{59}' \alpha _{51}''-\alpha _{24} \alpha _{91}' \alpha _{66}''+\alpha _{68} \alpha _{66}' \alpha _{91}''=0, 
\\
\alpha _{24} \alpha _{24}' \alpha _{22}''-\alpha _{51} \alpha _{51}' \alpha _{22}''+\left(\alpha _{22} \alpha _{11}'-\alpha _{11} \alpha _{22}'\right) \alpha _{24}''=0, \quad
\alpha _{24} \alpha _{24}' \alpha _{22}''-\alpha _{51} \alpha _{51}' \alpha _{22}''+\left(\alpha _{22} \alpha _{55}'-\alpha _{55} \alpha _{22}'\right) \alpha _{24}''=0, 
\\
\alpha _{22} \alpha _{68}' \alpha _{22}''+\alpha _{24} \alpha _{37}' \alpha _{24}''-\alpha _{33} \alpha _{68}' \alpha _{33}''-\alpha _{37} \alpha _{24}' \alpha _{37}''=0, \quad
\alpha _{37} \alpha _{37}' \alpha _{33}''-\alpha _{91} \alpha _{91}' \alpha _{33}''+\left(\alpha _{33} \alpha _{11}'-\alpha _{11} \alpha _{33}'\right) \alpha _{37}''=0, 
\\
\alpha _{91} \alpha _{51}' \alpha _{24}''+\alpha _{59} \left(\alpha _{22}' \alpha _{22}''-\alpha _{33}' \alpha _{33}''\right)-\alpha _{51} \alpha _{91}' \alpha _{37}''=0, \quad
\alpha _{37} \alpha _{37}' \alpha _{33}''-\alpha _{91} \alpha _{91}' \alpha _{33}''+\left(\alpha _{33} \alpha _{99}'-\alpha _{99} \alpha _{33}'\right) \alpha _{37}''=0, 
\\
\alpha _{22} \alpha _{51}' \alpha _{11}''-\alpha _{11} \alpha _{51}' \alpha _{22}''+\alpha _{22}' \left(\alpha _{24} \alpha _{51}''-\alpha _{51} \alpha _{24}''\right)=0, \quad
\alpha _{24} \left(\alpha _{22}' \alpha _{11}''-\alpha _{11}' \alpha _{22}''\right)+\alpha _{22} \left(\alpha _{51}' \alpha _{51}''-\alpha _{24}' \alpha _{24}''\right)=0, 
\\
\alpha _{22} \left(\alpha _{24}' \alpha _{24}''-\alpha _{51}' \alpha _{51}''\right)+\alpha _{24} \left(\alpha _{55}' \alpha _{22}''-\alpha _{22}' \alpha _{55}''\right)=0, \quad
\alpha _{55} \alpha _{51}' \alpha _{22}''+\alpha _{51} \alpha _{22}' \alpha _{24}''-\alpha _{24} \alpha _{22}' \alpha _{51}''-\alpha _{22} \alpha _{51}' \alpha _{55}''=0, 
\\
\alpha _{24} \alpha _{59}' \alpha _{37}''-\alpha _{91} \alpha _{11}' \alpha _{51}''-\alpha _{59} \alpha _{24}' \alpha _{55}''-\alpha _{99} \alpha _{37}' \alpha _{59}''=0, \quad
\alpha _{68} \alpha _{51}' \alpha _{37}''-\alpha _{11} \alpha _{37}' \alpha _{51}''-\alpha _{51} \alpha _{68}' \alpha _{55}''-\alpha _{91} \alpha _{99}' \alpha _{59}''=0, 
\\
\alpha _{22} \alpha _{37}' \alpha _{22}''+\alpha _{24} \alpha _{68}' \alpha _{24}''-\alpha _{66} \alpha _{37}' \alpha _{66}''-\alpha _{68} \alpha _{24}' \alpha _{68}''=0, \quad
\alpha _{33} \alpha _{24}' \alpha _{33}''+\alpha _{37} \alpha _{68}' \alpha _{37}''-\alpha _{66} \alpha _{24}' \alpha _{66}''-\alpha _{68} \alpha _{37}' \alpha _{68}''=0, 
\\
\alpha _{59} \alpha _{51}' \alpha _{24}''+\alpha _{91} \left(\alpha _{22}' \alpha _{22}''-\alpha _{66}' \alpha _{66}''\right)-\alpha _{51} \alpha _{59}' \alpha _{68}''=0, \quad
\alpha _{59} \alpha _{91}' \alpha _{37}''+\alpha _{51} \left(\alpha _{33}' \alpha _{33}''-\alpha _{66}' \alpha _{66}''\right)-\alpha _{91} \alpha _{59}' \alpha _{68}''=0, 
\\
\alpha _{59} \alpha _{59}' \alpha _{66}''-\alpha _{68} \alpha _{68}' \alpha _{66}''+\left(\alpha _{55} \alpha _{66}'-\alpha _{66} \alpha _{55}'\right) \alpha _{68}''=0, \quad
\alpha _{66} \alpha _{59}' \alpha _{55}''+\alpha _{68} \alpha _{66}' \alpha _{59}''-\alpha _{55} \alpha _{59}' \alpha _{66}''-\alpha _{59} \alpha _{66}' \alpha _{68}''=0, 
\\
\alpha _{68} \left(\alpha _{66}' \alpha _{55}''-\alpha _{55}' \alpha _{66}''\right)+\alpha _{66} \left(\alpha _{59}' \alpha _{59}''-\alpha _{68}' \alpha _{68}''\right)=0, \quad
\alpha _{59} \alpha _{59}' \alpha _{66}''-\alpha _{68} \alpha _{68}' \alpha _{66}''+\left(\alpha _{99} \alpha _{66}'-\alpha _{66} \alpha _{99}'\right) \alpha _{68}''=0, 
\\
\alpha _{33} \alpha _{91}' \alpha _{11}''-\alpha _{11} \alpha _{91}' \alpha _{33}''+\alpha _{33}' \left(\alpha _{37} \alpha _{91}''-\alpha _{91} \alpha _{37}''\right)=0, \quad
\alpha _{37} \alpha _{91}' \alpha _{51}''-\alpha _{22} \alpha _{22}' \alpha _{59}''+\alpha _{33} \alpha _{33}' \alpha _{59}''-\alpha _{24} \alpha _{51}' \alpha _{91}''=0, 
\\
\alpha _{33} \alpha _{33}' \alpha _{51}''-\alpha _{66} \alpha _{66}' \alpha _{51}''+\alpha _{37} \alpha _{91}' \alpha _{59}''-\alpha _{68} \alpha _{59}' \alpha _{91}''=0, \quad
\alpha _{68} \alpha _{59}' \alpha _{51}''-\alpha _{24} \alpha _{51}' \alpha _{59}''+\left(\alpha _{66} \alpha _{66}'-\alpha _{22} \alpha _{22}'\right) \alpha _{91}''=0, 
\\
\alpha _{91} \alpha _{24}' \alpha _{11}''+\alpha _{59} \alpha _{55}' \alpha _{51}''-\alpha _{24} \alpha _{91}' \alpha _{68}''+\alpha _{99} \alpha _{68}' \alpha _{91}''=0, \quad
\alpha _{37} \left(\alpha _{33}' \alpha _{11}''-\alpha _{11}' \alpha _{33}''\right)+\alpha _{33} \left(\alpha _{91}' \alpha _{91}''-\alpha _{37}' \alpha _{37}''\right)=0, 
\\
\alpha _{51} \alpha _{37}' \alpha _{11}''+\alpha _{55} \alpha _{68}' \alpha _{51}''-\alpha _{37} \alpha _{51}' \alpha _{68}''+\alpha _{59} \alpha _{99}' \alpha _{91}''=0, \quad
\alpha _{33} \left(\alpha _{37}' \alpha _{37}''-\alpha _{91}' \alpha _{91}''\right)+\alpha _{37} \left(\alpha _{99}' \alpha _{33}''-\alpha _{33}' \alpha _{99}''\right)=0, 
\\
\alpha _{37} \alpha _{59}' \alpha _{24}''-\alpha _{55} \alpha _{24}' \alpha _{59}''-\alpha _{51} \alpha _{11}' \alpha _{91}''-\alpha _{59} \alpha _{37}' \alpha _{99}''=0, \quad
\alpha _{68} \alpha _{66}' \alpha _{59}''-\alpha _{99} \alpha _{59}' \alpha _{66}''-\alpha _{59} \alpha _{66}' \alpha _{68}''+\alpha _{66} \alpha _{59}' \alpha _{99}''=0, 
\\
\alpha _{66} \left(\alpha _{59}' \alpha _{59}''-\alpha _{68}' \alpha _{68}''\right)+\alpha _{68} \left(\alpha _{66}' \alpha _{99}''-\alpha _{99}' \alpha _{66}''\right)=0, \quad 
\alpha _{68} \alpha _{91}' \alpha _{24}''-\alpha _{51} \alpha _{55}' \alpha _{59}''-\alpha _{11} \alpha _{24}' \alpha _{91}''-\alpha _{91} \alpha _{68}' \alpha _{99}''=0, 
\\
\alpha _{99} \alpha _{91}' \alpha _{33}''+\alpha _{91} \alpha _{33}' \alpha _{37}''-\alpha _{37} \alpha _{33}' \alpha _{91}''-\alpha _{33} \alpha _{91}' \alpha _{99}''=0 \\
\alpha _{22} \alpha _{24}' \alpha _{22}''+\alpha _{24} \alpha _{55}' \alpha _{24}''-\alpha _{51} \alpha _{11}' \alpha _{51}''-\alpha _{55} \alpha _{24}' \alpha _{55}''-\alpha _{59} \alpha _{37}' \alpha _{59}''=0, \\
\alpha _{24} \alpha _{51}' \alpha _{11}''-\alpha _{11} \alpha _{11}' \alpha _{51}''+\alpha _{22} \alpha _{22}' \alpha _{51}''-\alpha _{51} \alpha _{24}' \alpha _{55}''-\alpha _{91} \alpha _{37}' \alpha _{59}''=0 \\
\alpha _{55} \alpha _{51}' \alpha _{24}''-\alpha _{11} \alpha _{24}' \alpha _{51}''+\alpha _{51} \left(\alpha _{22}' \alpha _{22}''-\alpha _{55}' \alpha _{55}''\right)-\alpha _{91} \alpha _{68}' \alpha _{59}''=0, \\
-\alpha _{55} \alpha _{51}' \alpha _{24}''+\alpha _{11} \alpha _{24}' \alpha _{51}''+\alpha _{51} \left(\alpha _{55}' \alpha _{55}''-\alpha _{22}' \alpha _{22}''\right)+\alpha _{91} \alpha _{68}' \alpha _{59}''=0, \\
\alpha _{51} \alpha _{37}' \alpha _{51}''+\alpha _{55} \alpha _{68}' \alpha _{55}''+\alpha _{59} \alpha _{99}' \alpha _{59}''-\alpha _{66} \alpha _{68}' \alpha _{66}''-\alpha _{68} \alpha _{55}' \alpha _{68}''=0, \\
\alpha _{91} \alpha _{24}' \alpha _{51}''+\alpha _{99} \alpha _{68}' \alpha _{59}''+\alpha _{59} \left(\alpha _{55}' \alpha _{55}''-\alpha _{66}' \alpha _{66}''\right)-\alpha _{55} \alpha _{59}' \alpha _{68}''=0, \\
\alpha _{51} \left(\alpha _{11}' \alpha _{11}''-\alpha _{22}' \alpha _{22}''\right)-\alpha _{11} \alpha _{51}' \alpha _{24}''+\alpha _{55} \alpha _{24}' \alpha _{51}''+\alpha _{59} \alpha _{37}' \alpha _{91}''=0, \\
\alpha _{91} \left(\alpha _{11}' \alpha _{11}''-\alpha _{33}' \alpha _{33}''\right)-\alpha _{11} \alpha _{91}' \alpha _{37}''+\alpha _{59} \alpha _{24}' \alpha _{51}''+\alpha _{99} \alpha _{37}' \alpha _{91}''=0, \\
\alpha _{51} \alpha _{24}' \alpha _{11}''-\alpha _{22} \alpha _{22}' \alpha _{51}''+\alpha _{55} \alpha _{55}' \alpha _{51}''-\alpha _{24} \alpha _{51}' \alpha _{55}''+\alpha _{59} \alpha _{68}' \alpha _{91}''=0, \\
\alpha _{11} \alpha _{24}' \alpha _{11}''-\alpha _{22} \alpha _{24}' \alpha _{22}''-\alpha _{24} \alpha _{11}' \alpha _{24}''+\alpha _{51} \alpha _{55}' \alpha _{51}''+\alpha _{91} \alpha _{68}' \alpha _{91}''=0, \\
\alpha _{11} \alpha _{37}' \alpha _{11}''-\alpha _{33} \alpha _{37}' \alpha _{33}''-\alpha _{37} \alpha _{11}' \alpha _{37}''+\alpha _{51} \alpha _{68}' \alpha _{51}''+\alpha _{91} \alpha _{99}' \alpha _{91}''=0, \\
\alpha _{37} \alpha _{91}' \alpha _{11}''-\alpha _{51} \alpha _{24}' \alpha _{59}''-\alpha _{11} \alpha _{11}' \alpha _{91}''+\alpha _{33} \alpha _{33}' \alpha _{91}''-\alpha _{91} \alpha _{37}' \alpha _{99}''=0, \\
\alpha _{33} \alpha _{37}' \alpha _{33}''+\alpha _{37} \alpha _{99}' \alpha _{37}''-\alpha _{59} \alpha _{24}' \alpha _{59}''-\alpha _{91} \alpha _{11}' \alpha _{91}''-\alpha _{99} \alpha _{37}' \alpha _{99}''=0, \\
\alpha _{91} \alpha _{37}' \alpha _{51}''+\alpha _{59} \alpha _{68}' \alpha _{55}''-\alpha _{66} \alpha _{66}' \alpha _{59}''+\alpha _{99} \alpha _{99}' \alpha _{59}''-\alpha _{68} \alpha _{59}' \alpha _{99}''=0, \\
\alpha _{68} \alpha _{59}' \alpha _{55}''-\alpha _{55} \alpha _{55}' \alpha _{59}''+\alpha _{66} \alpha _{66}' \alpha _{59}''-\alpha _{51} \alpha _{24}' \alpha _{91}''-\alpha _{59} \alpha _{68}' \alpha _{99}''=0, \\
\alpha _{59} \alpha _{55}' \alpha _{59}''-\alpha _{66} \alpha _{68}' \alpha _{66}''-\alpha _{68} \alpha _{99}' \alpha _{68}''+\alpha _{91} \alpha _{24}' \alpha _{91}''+\alpha _{99} \alpha _{68}' \alpha _{99}''=0, \\
\alpha _{91} \alpha _{37}' \alpha _{11}''+\alpha _{59} \alpha _{68}' \alpha _{51}''-\alpha _{33} \alpha _{33}' \alpha _{91}''+\alpha _{99} \alpha _{99}' \alpha _{91}''-\alpha _{37} \alpha _{91}' \alpha _{99}''=0, \\
\alpha _{55} \alpha _{68}' \alpha _{59}''-\alpha _{99} \alpha _{59}' \alpha _{68}''+\alpha _{51} \alpha _{37}' \alpha _{91}''+\alpha _{59} \left(\alpha _{99}' \alpha _{99}''-\alpha _{66}' \alpha _{66}''\right)=0, \\
\alpha _{99} \alpha _{91}' \alpha _{37}''-\alpha _{51} \alpha _{68}' \alpha _{59}''-\alpha _{11} \alpha _{37}' \alpha _{91}''+\alpha _{91} \left(\alpha _{33}' \alpha _{33}''-\alpha _{99}' \alpha _{99}''\right)=0, \\
-\alpha _{99} \alpha _{91}' \alpha _{37}''+\alpha _{51} \alpha _{68}' \alpha _{59}''+\alpha _{11} \alpha _{37}' \alpha _{91}''+\alpha _{91} \left(\alpha _{99}' \alpha _{99}''-\alpha _{33}' \alpha _{33}''\right)=0. 
\end{align}
\end{small}

\bibliographystyle{elsarticle-num}

\bibliography{su_n}
\end{document}